 \documentclass[twocolumn, epjc3]{svjour3}
 \usepackage{graphicx, amsmath, amstext, amssymb, amsfonts,amsmath}
 \RequirePackage[colorlinks=true,citecolor=blue,urlcolor=blue,linkcolor=blue]{hyperref}
 \RequirePackage[T1]{fontenc}
 \smartqed                  
 \RequirePackage{mathptmx}  
 \RequirePackage{flushend}
 \RequirePackage[numbers,sort&compress]{natbib}
\journalname{Eur. Phys. J. C}
\begin{document}
  \title{Black hole thermodynamical entropy}
  \author{Constantino Tsallis\thanksref{addr1, addr2, email1} \and Leonardo J.L. Cirto\thanksref{addr1, email2}}
  \institute{Centro Brasileiro de Pesquisas Fisicas and National Institute of Science and Technology for Complex Systems, Rua Xavier Sigaud 150, 22290-180 Rio de Janeiro-RJ, Brazil\label{addr1} \and Santa Fe Institute, 1399 Hyde Park Road, Santa Fe, NM 87501, USA\label{addr2}}
  \thankstext{email1}{e-mail: tsallis@cbpf.br}
  \thankstext{email2}{e-mail: cirto@cbpf.br}
  \date{Received: date / Accepted: date}
  \maketitle
\begin{abstract}
As early as 1902, Gibbs pointed out that systems whose partition function diverges, e.g. gravitation, lie outside the validity of the Boltzmann-Gibbs (BG) theory.
Consistently, since the pioneering Bekenstein-Hawking results, physically meaningful evidence (e.g., the holographic principle) has accumulated that the BG entropy $S_{BG}$  of a $(3+1)$ black hole is
proportional to its area $L^2$ ($L$ being a characteristic linear length), and not to its volume $L^3$. Similarly it exists the \emph{area law}, so named because, for a wide class of strongly
quantum-entangled $d$-dimensional systems, $S_{BG}$ is proportional to $\ln L$ if $d=1$, and to $L^{d-1}$ if $d>1$, instead of being proportional to $L^d$ ($d \ge 1$).
These results violate the extensivity of the thermodynamical entropy of a $d$-dimensional system.
This thermodynamical inconsistency disappears if we realize that the thermodynamical entropy of such nonstandard systems is \emph{not} to be identified with the BG {\it additive} entropy  but with appropriately generalized {\it nonadditive} entropies. Indeed, the celebrated usefulness of the BG entropy is founded on hypothesis such as relatively weak probabilistic correlations (and their connections to ergodicity, which by no means can be assumed as a general rule of nature). Here we introduce a generalized entropy which, for the
Schwarzschild black hole and the area law, can solve the thermodynamic puzzle.
\end{abstract}
\section{Introduction}
The entropy of a black hole presents intriguing aspects that are being currently discussed since several decades.
Indeed, since the pioneering works of Bekenstein~\cite{Bekenstein1973, Bekenstein1973_2} and Hawking~\cite{Hawking1974, Hawking1974_2},  it has become frequent in the literature the (explicit or tacit) acceptance that the black-hole entropy is ano\-ma\-lous in the sense that it violates thermodynamical extensivity~\cite{tHooft1985,tHooft1990,Susskind1993,Maddox1993,Srednicki1993,StromingerVafa1996,MaldacenaStrominger1998,DasandShankaranarayanan2006, BrusteinEinhornYarom2006, BorstenDahanayakeDuffEbrahimRubens2009,Padmanabhan2009,Casini2009,BorstenDahanayakeDuffMarraniRubens2010,Corda2011,KolekarPadmanabhan2011,Saida2011}.
We frequently read and hear claims that the entropy of the black hole is proportional to the area of its boundary instead of being proportional to the black-hole volume.
To discuss this interesting question within  a thermodynamically proper context, let us first remind a typical form of the thermodynamical energy $G$ of a generic $d$-dimensional system~\cite{Callen}:
\begin{eqnarray}
\begin{split}
&G(V,T,p,\mu,H, \dots) =  U(V,T,p,\mu,H,\dots) + \\ 
& - TS(V,T,p,\mu,H,\dots) +pV -\mu N(V,T,p,\mu,H,\dots)+ \\
& -HM(V,T,p,\mu,H,\dots)- \cdots \,,
\label{thermodynamics}
\end{split}
\end{eqnarray}
where $T, p, \mu,H$ are the temperature, pressure, chemical potential, external magnetic field, and $U,S,V,N,M$ are the internal energy, entropy, volume, number of particles (in turn proportional to the number of degrees of freedom), magnetization. We identify three classes of variables, namely (i) those that are expected to always be extensive ($S,V, N,M,\ldots$), i.e., scaling with $V=L^d$, where $L$ is a characteristic linear dimension of the system (clearly, $V \propto A^{d/(d-1)}$, where $A$ is the area), (ii) those that characterize the external conditions under which the system is placed ($T,p,\mu,H,\ldots$), scaling with $L^\theta$, and (iii) those that represent energies ($G,U$), scaling with $L^\epsilon$.
It trivially follows
\begin{equation}
\epsilon = \theta + d \,.
\label{thetad}
\end{equation}
If we divide Eq. (\ref{thermodynamics}) by $L^{\theta + d}$ and consider the large $L$ limit (i.e., the thermodynamical limit), we obtain
 \begin{eqnarray}
\begin{split}
& g \Bigl(\frac{T}{L^\theta},\frac{p}{L^\theta},\frac{\mu}{L^\theta},\frac{H}{L^\theta}, \dots \Bigr) =
  u\Bigl(\frac{T}{L^\theta},\frac{p}{L^\theta},\frac{\mu}{L^\theta},\frac{H}{L^\theta}, \dots \Bigr) + \\
& -\frac{T}{L^\theta}\,s\Bigl(\frac{T}{L^\theta},\frac{p}{L^\theta},\frac{\mu}{L^\theta},\frac{H}{L^\theta}, \dots \Bigr) 
  +\frac{p}{L^\theta} + \\
& -\frac{\mu}{L^\theta} \,n\Bigl(\frac{T}{L^\theta},\frac{p}{L^\theta},\frac{\mu}{L^\theta},\frac{H}{L^\theta}, \dots \Bigr) 
 -\frac{H}{L^\theta}\,m\Bigl(\frac{T}{L^\theta},\frac{p}{L^\theta},\frac{\mu}{L^\theta},\frac{H}{L^\theta}, \dots\Bigr)- \cdots \,,
\label{thermodynamics2}
\end{split}
\end{eqnarray}
where  $g \equiv \lim_{L\to\infty} G/L^{\theta + d}$,  $u \equiv \lim_{L\to\infty} U/L^{\theta +d}$, $s \equiv \lim_{L\to\infty} S/L^d$, $n \equiv \lim_{L\to\infty} N/L^d$, $m \equiv \lim_{L\to\infty} M/L^d$.
The correctness of all the scalings appearing in this equation has been profusely verified in the literature for (both short- and long-range interacting) thermal~\cite{thermal, thermal_2, thermal_3, thermal_4, thermal_5, thermal_6, thermal_7}, diffusion~\cite{diffusion} and geometrical (percolation) systems~\cite{geometrical, geometrical_2}.

Next, let us illustrate relation (\ref{thetad}) through four different physical situations. First, for a standard thermodynamical system (e.g., a real gas, a simple metal) we have $\theta =0$ (i.e., usual intensive variables), and $\epsilon = d$ (i.e., usual extensive variables). This is the answer that is found in the textbooks of thermodynamics.

Second, for a classical many-body Hamiltonian system with two-body long-range (attractive) interactions asymptotically decaying with distance $r$ like $1/r^\alpha \;\;(0 \le \alpha < d)$ we have indeed \cite{thermal, thermal_2, thermal_3, thermal_4, thermal_5, thermal_6, thermal_7, diffusion,geometrical, geometrical_2,Tsallis2009} $\theta =d-\alpha$, hence, using relation (\ref{thetad}), $\epsilon=2d-\alpha$. These peculiar scalings are a consequence from the fact that such potential is not integrable, i.e., from the fact that the integral $\int_{\textrm{constant}}^\infty dr \,r^{d-1}\,r^{-\alpha}$ diverges, and therefore the Boltzmann-Gibbs (BG) canonical partition function itself diverges. In his 1902 book {\it Elementary Principles in Statistical Mechanics} \cite{Gibbs1902},  Gibbs himself emphatically points out that whenever the partition function diverges, the BG theory can not be used (in his words ``the law of distribution becomes illusory").
As an illustration of his remark he refers specifically to the case of Newtonian gravitation~\cite{remark}.

Third, for a Schwarzschild $(3+1)$-dimensional black hole, the energy scales like the mass $M_{bh}$ (where bh stands for {\it black hole}), which in turn scales with $L$ \cite{generalrelativity, generalrelativity_2, generalrelativity_3}, hence $\epsilon =1$, hence, using Eq.~(\ref{thetad}),
\begin{equation}
\theta = 1-d\,.
\label{1d}
\end{equation}
{\it If} the black hole is physically identified with its event horizon surface, then it is to be considered as a genuine $d=2$ system, then $\theta =-1$, which precisely recovers the usual Bekenstein-Hawking (BH) scaling $T \propto 1/L \propto 1/M_{bh}$.
\emph{If} however the black hole is to be considered as a genuine $d=3$ system (which makes sense given that the corresponding space-time is (3+1)-dimensional), then $\theta =-2$, i.e., $T$ scales like $1/L^2 \propto 1/M_{bh}^2$, in variance with the BH scaling. This is a manner for understanding why such a puzzle exists since decades related to the entropy of a black hole. Let us be somewhat more specific. Wide and physically meaningful evidence (e.g., the holographic principle) exists that the {\it Boltzmann-Gibbs} entropy (for quantum systems, also referred to as {\it von Neumann} entropy) $S_{BG} \equiv k_B\ln W \propto L^2$, and more generally that $S_{BG} \equiv -k_B Tr \rho \ln \rho \propto L^2$, $W$ being the total number of internal configurations,
and $\rho$ being the density matrix.
For strongly quantum-entangled $d$-dimensional systems we similarly have what is currently referred to as the {\it area law} \cite{EisertCramerPlenio2010}, i.e., the fact that $S_{BG} \equiv -k_BTr \rho \ln \rho$ frequently scales with $L^{d-1}$ for $d>1$, and with $\ln L$ for $d=1$, instead of scaling, for $d \ge 1$, with $L^d$. This fact also generates a closely related intriguing question. The above remarks might be considered the heart of the ongoing discussion for the entropy of a black hole. Indeed, if the system is to be physically considered a $(d-1)$-dimensional one, then the (additive) entropy $S_{BG}$ certainly is to be identified as its thermodynamical entropy. But if the system is to be physically considered a $d$-dimensional one, then $S_{BG}$ can {\it not} be identified as its thermodynamical entropy, and, as we shall soon see, a nonadditive entropic functional is needed  to play that role.

Fourth, a $(2+1)$-dimensional ``black hole" has been discussed as well~\cite{carlip, carlip_2, carlip_3}.
It has been shown that the energy scales like~$L^2$, hence $\epsilon=2$ and, using Eq.~\eqref{thetad} once again, 
\begin{equation}
\theta=2-d \,.
\end{equation}
This case provides an event horizon which is one-dimen\-sional.
If, due to this fact, this black hole is to be considered a genuine $d=1$ system,
then $\theta = 1$,  which corresponds to the $(2+1)$ version of BH scaling, i.e., $T\propto L\propto M_{bh}^{1/2}$.
Indeed, this is precisely the scaling that has been obtained~\cite{carlip, carlip_2, carlip_3} for this simplified system.
If, however, this black hole is to be considered as a $d=2$ system, we have $\theta=0$, and, in this case, $T$ is expected to be an intensive variable.
Consistently, if we assume the system to be a $d=1$ one, then clearly $S_{BG}$ plays the role of its thermodynamical entropy, since~$S_{BG}\propto L$ (as obtained in~\cite{carlip, carlip_2, carlip_3}).
But, similarly to the $(3+1)$ case discussed above, if we consider it to be a $d=2$ one, then once again a nonadditive entropic functional is needed to play the thermodynamical role.

The physical system we primarily focus on in the present paper is a~$(3+1)$ black hole like that of Schwarzschild. As emphasized above, if we are to consider it as a genuine~$d=2$ system, then $S_{BG} = k_B \ln W \propto L^2$ corresponds indeed to the (extensive) thermodynamical entropy $S$, the BH scaling $T \propto 1/M_{bh}$ is to be expected, and there are no controversial or intriguing facts to be further analyzed.
If however, this black hole is to be considered a genuine~$d=3$ system, then~$S_{BG}$ {\it can not be the thermodynamical entropy~$S$}, since the latter must scale like~$L^3$ while $S_{BG}$ scales like $L^2$. Within this standpoint, a crucial question then arises, namely,  {\it what is then the microscopic mathematical expression of the thermodynamical entropy $S$ of this 3-dimensional system?}
The purpose of the present paper is to provide a thermodynamically admissible answer to this important question.

From a historical perspective, we observe that, strangely enough, Gibbs's crucial remark and the dramatic theoretical features to which it is related are often overlooked in textbooks.
Similarly, the thermodynamical violation related to the area law frequently is, somehow, not taken that seriously. Indeed, not few authors seem inclined to consider that, for such complex systems, the entropy is not expected to satisfy thermodynamical extensivity.
However, various physical and mathematical facts exist which reveal such standpoint as kind of bizarre.
The specific goal of the present paper is to address this important issue and develop a path along which the difficulty can be overcome.
The fact (repeatedly illustrated in various manners for strongly entangled systems, black holes and, generically speaking, for systems satisfying the above mentioned area law) that the Boltzmann\--Gibbs\--von Neumann (additive) entropy is {\it not} proportional to the volume $L^d$ precisely shows that, for such strongly correlated systems
(hence the total number of admissible states in phase space is sensibly reduced), {\it the thermodynamical entropy can not be identified with the usual one but with a substantially different (nonadditive) one}.

An argument reinforcing the correctness of using nonadditive entropic forms in order to re-establish the entropic extensivity of the system can be found in the results achieved by Hanel and Thurner~\cite{HanelThurner2011, HanelThurner2011_2} by focusing on the Khinchine axioms and on complex systems with surface-dominant sta\-tistics.

A further indication we can refer to is the analogy with the time $t$ dependence of the entropy of simple nonlinear dynamical systems, e.g., the logistic map.
Indeed, for the parameter values for which the system has positive Lyapunov exponent (i.e., strong chaos and ergodicity), we verify~$S_{BG}~\propto~t$ (under appropriate mathematical limits), but for parameter values where the Lyapunov exponent vanishes (i.e., weak chaos and breakdown of ergodicity), it is a nonadditive entropy ($S_q$, discussed below) the one which grows linearly with $t$ (see~\cite{BaldovinRobledo2004, BaldovinRobledo2004_2, BaldovinRobledo2004_3, BaldovinRobledo2004_4, BaldovinRobledo2004_5, BaldovinRobledo2004_6, BaldovinRobledo2004_7, BaldovinRobledo2004_8, BaldovinRobledo2004_9, BaldovinRobledo2004_10} and references therein), and consistently provides a generalized Pesin-like identity.
If we take into account that, in many such dynamical systems, $t$ plays a role analogous to $N$ in thermodynamical systems, we have here one more indication which aligns with the extensivity of the entropy for complex systems.

Finally, one more recent result exists~\cite{RuizTsallis2012, RuizTsallis2012_2, RuizTsallis2012_3}, related to the so called Large Deviation Theory in theory of probabilities, which also is consistent with the extensivity of the entropy, even in the presence of strong correlations between the elements of the system.
\section{Nonadditive entropies}
Let us now turn onto the fact that entropies generalizing that of BG become necessary in order to recover thermodynamic extensivity for nonstandard systems.
Let us first describe briefly an entropic functional form, $S_q$, generalizing that of BG, which has been successfully applied for many complex systems, as illustrated below.
After that we shall address another such generalization, $S_\delta$ (see Eq.~\ref{deltaentropy}), which constitutes in fact one of the contributions of the present work.

As a possibility for addressing complexities such as tho\-se illustrated above, it was proposed in 1988~\cite{Tsallis1988} (see also \cite{GellMannTsallis2004,TsallisGellMannSato2005,Tsallis2009}) a generalization of the BG theory, currently referred to as nonextensive statistical mechanics.
It is based on the nonadditive entropy
\begin{equation}
S_q=k_B\frac{1-\sum_{i=1}^Wp_i^q}{q-1}
= k_B  \sum_{i=1}^W p_i \ln_q \frac{1}{p_i} \;\; \left(q \in {\cal R}; \, \sum_{i=1}^W p_i=1 \right),
\label{qentropy}
\end{equation}
with $\ln_q z \equiv (z^{1-q}-1)/(1-q)$  ($\ln_1 z=\ln z$).
$S_q$ recovers $S_{BG}= -k_B\sum_{i=1}^W p_i \ln p_i$ for $q\to 1$.
If $A$ and $B$ are two {\it probabilistically independent} systems (i.e., $p_{ij}^{A+B}=p_i^Ap_j^B$, $\forall (i,j)$), definition (\ref{qentropy}) implies
\begin{equation}
\frac{S_q(A+B)}{k_B} =  \frac{S_q(A)}{k_B}+ \frac{S_q(B)}{k_B}
+ (1-q)\frac{S_q(A)}{k_B}\frac{S_q(B)}{k_B} \,.
\end{equation}
In other words, according to the definition of entropic additivity in~\cite{Penrose1970},  $S_q$ is additive if $q=1$, and nonadditive otherwise.

If probabilities are all equal, we straightforwardly obtain
\begin{equation}
S_q=k_B \ln_q W \,.
\end{equation}
If we extremize (\ref{qentropy}) with a (finite) constraint on the width of the probability distribution $\{p_i\}$ (in addition to its normalization), we obtain
\begin{equation}
p_i=\frac{e_q^{-\beta_q\,E_i}}{\sum_{j=1}^W e_q^{-\beta_q\,E_j}} \,,
\label{pq}
\end{equation}
$e_q^z$ being the inverse of the $q$-logarithmic function, i.e., $e_q^z \equiv [1+(1-q)z]^{1/(1-q)}$ ($e_1^z=e^z$); $\{E_i\}$ are the energy levels; $\beta_q$ is an effective inverse temperature.

Complexity frequently emerges in natural, artificial and social systems. It may be caused by various geometrical-dynamical ingredients, which include non-ergodicity, long-term memory, multifractality, and other spatial-temporal long-range correlations between the elements of the system.
During the last two decades, many such phenomena have been successfully approached in the frame of nonextensive statistical mechanics.
Predictions, verifications and various applications have been performed in high-energy physics~\cite{CMS1_0, Wong_Wilk_PRD_2013, CMS1, CMS1_2, ALICE1_3, ALICE1_3_2, ATLAS, PHENIX, PHENIX_2, ShaoYiTangChenLiXu2010},  spin-glasses \cite{PickupCywinskiPappasFaragoFouquet2009},  flux of cosmic rays \cite{TsallisAnjosBorges2003}, turbulence in pure-electron plasma \cite{AnteneodoTsallis1997}, self-organized criticality in biological evolution \cite{TamaritCannasTsallis1998}, cold atoms in optical lattices \cite{DouglasBergaminiRenzoni2006},  trapped ions \cite{DeVoe2009},  anomalous diffusion \cite{AndradeSilvaMoreiraNobreCurado2010, AndradeSilvaMoreiraNobreCurado2010_2, AndradeSilvaMoreiraNobreCurado2010_3, AndradeSilvaMoreiraNobreCurado2010_4, AndradeSilvaMoreiraNobreCurado2010_5, AndradeSilvaMoreiraNobreCurado2010_6},  dusty plasmas~\cite{LiuGoree2008},   solar physics~\cite{BurlagaVinasNessAcuna2006, BurlagaVinasNessAcuna2006_2, BurlagaVinasNessAcuna2006_3},
relativistic and nonrelativistic nonlinear quantum mechanics~\cite{NobreMonteiroTsallis2011, NobreMonteiroTsallis2011_2, NobreMonteiroTsallis2011_3},  among many others.

If a physical system is constituted by $N$ elements, and these elements are independent (or quasi-independent in so\-me sen\-se), we have that
\begin{equation}
W(N) \sim A \xi^N \;\;\; (A>0;\, \xi>1; \, N\to\infty) \,.
\label{A}
\end{equation}
(For example, for $N$ independent coins, we have $W=2^N$.)
Therefore, by illustrating the present point for the particular case of equal probabilities, we immediately verify that
\begin{equation}
S_{BG}(N) = k_B \ln W(N) \sim k_B (\ln \xi) N \propto N \;\;\;(N\to\infty)\,,
\end{equation}
hence thermodynamical extensivity is satisfied. This reconfirms that, for such systems, the thermodynamically admissible
entropy is precisely given by the additive one, $S_{BG}$, as well known. If, however, strong correlations are present (of the type assumed in the $q$-generalization of the Central Limit and L\'evy-Gnedenko Theorems~\cite{UmarovTsallisSteinberg2008, UmarovTsallisSteinberg2008_2}), we can have
\begin{equation}
W(N) \sim B N^\tau \;\;\; (B>0; \, \tau>0; \, N \to\infty) \,.
\label{B}
\end{equation}
In this case, we straightforwardly verify that, for $q=1-\frac{1}{\tau}$,
\begin{equation}
S_{q}(N) = k_B \ln_q W(N) \propto N \;\;\;(N\to\infty)\,,
\end{equation}
which satisfies thermodynamical extensivity, in contrast with $S_{BG}(N) \propto \ln N$, which violates it.
Probabilistic and physical models which belong to this class are respectively available in~\cite{TsallisGellMannSato2005} and~\cite{CarusoTsallis2008, SaguiaSarandy2010}.

\begin{figure}[t] 
\begin{center}
  \includegraphics[width=1.0\linewidth]{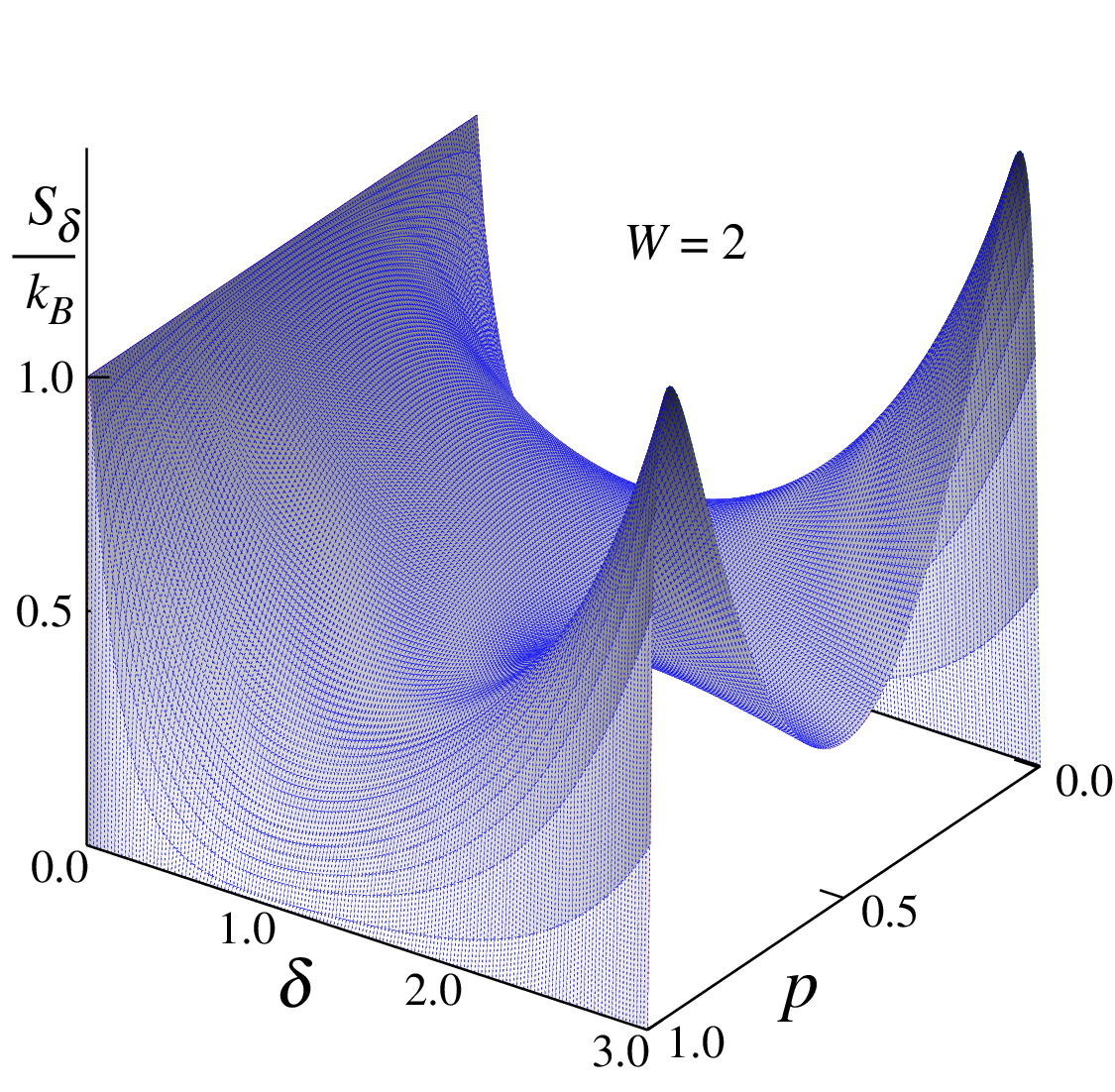}
\end{center}   
\caption
{
Entropy $S_{\delta}$ as a function of the index $\delta$ and the probability $p$ of a binary variable ($W=2$).
Concavity is lost for $\delta > 1 + \ln 2$.
}
\label{figSdelta_3D}
\end{figure}
It is clear that, for $N\gg 1$, expression (\ref{B}) becomes increasingly smaller than (\ref{A}).

However there are cases which are described neither by Eq. (\ref{A}) nor by Eq. (\ref{B}). Such is the case if
\begin{equation}
W(N) \sim C  \nu^{N^\gamma}\;\;\; (C>0\, ; \nu > 1; \, 0 < \gamma < 1) \,,
\label{C}
\end{equation}
which also becomes increasingly smaller that (\ref{A}) (though larger than (\ref{B})).
More precisely, we have
\begin{eqnarray}
\begin{split}
B N^\tau \ll C  \nu^{N^\gamma} \ll A \xi^N 
\,\,\,(N \gg 1;\, \,A>0; \, B>0; \\ \,C>0;\,  \xi >1;\, \nu > 1; \, \tau >0; \, 0 < \gamma < 1) \,.
\end{split}
\end{eqnarray}

\begin{figure}
\begin{center}
\includegraphics[width=1.0\linewidth]{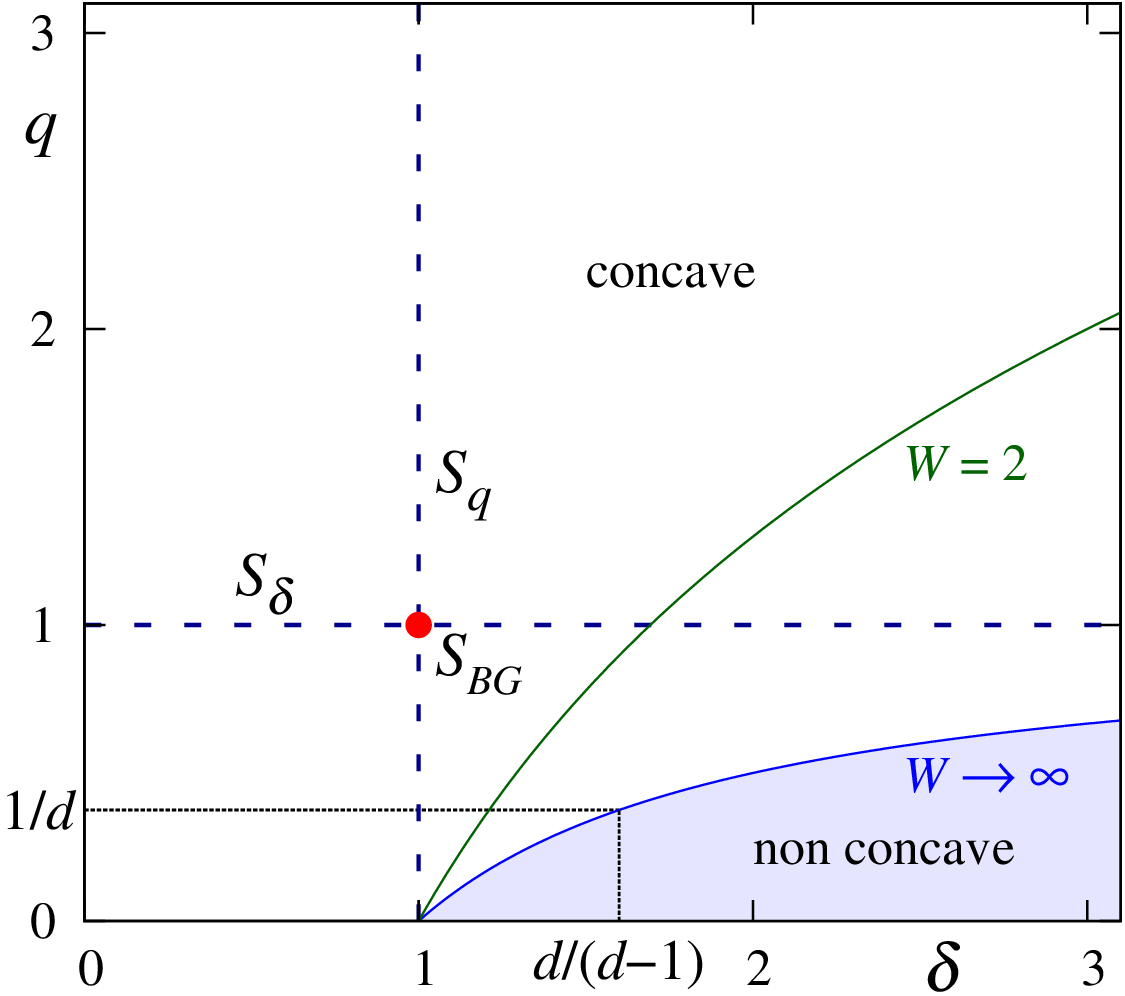}
\end{center}
\caption{Parameter space $(q,\delta)$ of the entropy $S_{q,\delta}$.
At the point $(1,1)$ we recover the Boltzmann-Gibbs entropy $S_{BG}$.
At $\delta =1$ ($q=1$) we recover the nonadditive entropy $S_q$ ($S_\delta$).  For any fixed $W$ there is a frontier  $q(\delta)$ such that, for $\delta$ values at its left, the entropy $S_{q,\delta}$ is concave, and, at its right, it neither concave nor convex.
The $W=2$ and $W \to\infty$ frontiers are indicated in the plot.
The entropy $S_\delta$ is concave for $0<\delta\leq1+\ln W$.
If we impose the extensivity of $S_{q,\delta}$ for the class of systems represented by Eq.~(\ref{stretched}), it must be $\delta = 1/\gamma \ge 1$. If $S_{q,\delta}$ is used for other purposes, the region $0<\delta<1$ is accessible as well.}
\label{figqdelta} 
\end{figure}

The entropy associated with $\gamma \to 1$ is of course $S_{BG}$. What about $0< \gamma <1$ ?
A simple answer is in fact already available in the literature (footnote of page 69 in~\cite{Tsallis2009}), namely,
\begin{equation}
S_\delta=k_B \sum_{i=1}^W p_i \left(\ln\frac{1}{p_i} \right)^\delta \;\;\; (\delta > 0) \,.
\label{deltaentropy}
\end{equation}
The case $\delta=1$ recovers $S_{BG}$.
This entropy is, like $S_q$ for $q>0$, concave for $0< \delta \le (1+ \ln W$).
And, also like $S_q$ for $q \ne 1$, it is nonadditive for $\delta \ne 1$.
Indeed, for probabilistically independent systems $A$ and $B$ (hence $W^{A+B}=W^AW^B$), we verify $S_{\delta}(A+B)\neq S_{\delta}(A)+S_{\delta}(B)$.
For equal probabilities we have
\begin{equation}
S_\delta = k_B \ln^\delta W \,,
\end{equation}
hence, for $\delta >0$,
\begin{equation}
\frac{S_\delta(A+B)}{k_B} =  \left\{   \left[ \frac{S_\delta(A)}{k_B} \right]^{1/\delta} +     \left[  \frac{S_\delta(B)}{k_B} \right]^{1/\delta}   \right\}^\delta \,.
\label{deltaentropy2}
\end{equation}
It is easily verified that, if $W(N)$ satisfies (\ref{C}), $S_\delta(N)$ is extensive for $\delta=1/\gamma$. This is in fact true even if
\begin{equation}
W(N) \sim \phi(N)\nu^{N^\gamma} \;\;(\nu>1; \, 0<\gamma<1) \,,
\label{stretched}
\end{equation}
$\phi(N)$ being any function satisfying $\lim_{N\to\infty} \frac{\ln \phi(N)}{N^\gamma}=0$, for example $\phi(N)=BN^\tau$.
 Let us now unify $S_q$ (Eq.~(\ref{qentropy})) and $S_\delta$ (Eq.~(\ref{deltaentropy})) as follows:
\begin{eqnarray}
S_{q,\delta}
= k_B  \sum_{i=1}^W p_i \left(\ln_q \frac{1}{p_i}\right)^\delta \,.
\label{qdeltaentropy}
\end{eqnarray}
$S_{q,1}$ and $S_{1,\delta}$ respectively recover $S_q$ and $S_\delta$; $S_{1,1}$ recovers $S_{BG}$. Obviously this entropy is nonadditive unless $(q,\delta)=(1,1)$, and it is expansible (see, for instance, \cite{Tsallis2009}), $\forall q>0$, $\forall \delta>0$.
It is concave for all $q>0$ and $0<\delta \le (qW^{q-1}-1)/(q-1)$.
In the limit $W \to\infty$, this condition becomes $0<\delta \le 1/(1-q), \,\forall q \in(0,1)$, and any $\delta>0$ for $q\ge 1$; see Figs.~\ref{figSdelta_3D} and~\ref{figqdelta}. For equal probabilities we have
\begin{equation}
S_{q,\delta}=k_B (\ln_q W)^\delta \,.
\end{equation}

The above results for the equal-probabilities case may be summarized as follows. If we have
\begin{equation}
W(N) \sim B N^\tau \nu^{N^\gamma} \;\;(B>0; \, \tau \ge 0; \,\nu>1; \, 0 \le \gamma \le 1) \,,
\label{stretched2}
\end{equation}
$S_{q,\delta}$ is extensive (i.e., $S_{q,\delta} \propto N$, for $N \to\infty$) for $(q,\delta)=(1,1)$ if $(\gamma,\tau)=(1,0)$ (notice that $\tau >0$ is inadmissible if $\gamma =1$, since no occupancy of phase space can be larger than full occupancy), for $(q,\delta)=(1-1/\tau,1)$ if $\gamma =0$ and $\tau >0$, and for
 $(q,\delta)=(1,1/\gamma)$ if $0< \gamma <1$.

 Let us  mention at this point that several two-parameter entropic functionals different from $S_{q,\delta}$ are in fact available in the literature
(see, for instance, \cite{BorgesRoditi1998, BorgesRoditi1998_2, HanelThurner2011, HanelThurner2011_2};  see also~\cite{Tempesta2011}).
In particular the asymptotic behaviours of $S_{q,\delta}$ in the thermodynamic limit coincide, for all values of $(q,\delta)$, with those of the recently introduced Hanel-Thurner entropy $S_{c,d}$ \cite{HanelThurner2011, HanelThurner2011_2} for appropriate values of~$(c,d)$.

\section{Discussion and conclusion}
We can address now the area law.
It has been verified for those anomalous $d$-dimensional systems that essentially yield $\ln W(L) \propto L^{d-1}$ ($d>1$), which implies that $W(L)$ is of the type indicated in (\ref{stretched}), more precisely that
\begin{equation}
W(L) \sim \phi(L^d) \nu^{L^{d-1}} = \phi(L^d) \nu^{{(L^d)}^{(d-1)/d}}
\end{equation}
Therefore, $S_\delta =S_{1,\delta}$ for $\delta = d/(d-1)$ is extensive, thus satisfying thermodynamics. For equal probabilities, we straightforwardly verify that
\begin{equation}
\frac{S_{\delta=d/(d-1)}}{k_B} \propto \left(\frac{ S_{BG}}{k_B}\right)^{d/(d-1)} \hspace{1cm} (d>1).
\end{equation}
Moreover, for such anomalous systems, the entropy $S_{\delta=d/(d-1)}$ is expected to be extensive for arbitrary density matrices, and not only for the simple equal-probability case.
For $d=~3$, it can be connected with the well-known Bekenstein-Hawking entropy $S_{BH}$ through
\begin{equation}
\frac{S_{\delta=3/2}}{k_B} \propto \left(\frac{ S_{BH}}{k_B}\right)^{3/2} \,,
\label{eq:qdeltaS_E_BHS}
\end{equation}
where
\begin{equation}
S_{BH} = \frac{k_B}{4} \frac{A_H}{G\hbar/c^3}\,,
\end{equation}
$A_H$ being the event horizon area.
It is important to stress that Eq.~\eqref{eq:qdeltaS_E_BHS} has \emph{not} been imposed in an \emph{ad hoc} manner just to transform $L^{d-1}$ into $L^d$\,: it has been derived from a new entropic functional, namely $S_{\delta}$.
This entropy $S_{\delta}$ has been defined under the assumption that the current black-hole result $\ln W \propto A_{H}$ is correct.
Also, by using the fact that~$d/(d-1)~>~0$, we verify that $S_{\delta=d/(d-1)}$ increases monotonically with $S_{BH}$.
This is consistent with the second principle of thermodynamics, namely that whenever $S_{BH}$ increases with time, so does $S_{\delta=3/2}$ (and the same happens in general with  $S_{\delta=d/(d-1)}$).

At the present state of knowledge we cannot exclude the possibility of extensivity of $S_{q,\delta}$ for other special values of~$(q,\delta)$, particularly in the limit $\delta \to\infty$.

For a block of a~$d=1$ gapless fermionic system, it has been analytically proved~\cite{CarusoTsallis2008} the extensivity of $S_q$ for a specific value of $q<1$ which depends on the central charge of the universality class that is being focused on (see also~\cite{SaguiaSarandy2010} for a different type of $d=1$ system).
For a~$d=2$ gapless bosonic system,  it has been numerically found~\cite{CarusoTsallis2008} that, once again, it is $S_q$ (with a value of $q<1$) the entropy which is extensive and consequently satisfies thermodynamics. This kind of scenario might be present in many $d$-dimensional physical systems for which $\ln W(N) \propto \ln_{2-d} N$ (i.e., $\propto \ln L$ for $d=1$, and $\propto L^{d-1}$ for $d>1$).

Summarizing, classical thermodynamics and the thermostatistics of a wide class of systems whose elements are strongly correlated
(for instance, through long-range interactions, or through strong quantum entanglement, or both, such as black holes, quantum gravitational dense systems, and others)
can be reconciled (along lines similar to those illustrated  in~\cite{GellMannTsallis2004,TsallisGellMannSato2005,CarusoTsallis2008} for simple cases).
It is enough, for such complex systems, to identify the thermodynamical entropy with nonadditive entropies such as $S_{q, \delta}$, and not necessarily with the usual Boltzmann-Gibbs-von Neumann one, which corresponds to $q=\delta=1$. This statement is by no means in conflict with the well accepted relation that, for a Schwarzschild $(3+1)$-dimensional black hole,  $S_{BG}\propto area$.

\begin{acknowledgements}
We acknowledge  useful conversations with M. Jauregui.
One of us (CT) also acknowledges (recent and old) conversations with L. Bergstrom, F. Caruso, H. Casini, A. Coniglio, E.M.F. Curado, M.J. Duff, A.S. Fokas, G. 't~Hooft, F.D. Nobre, N. Pinto Neto, G. Ruiz,  H. Saida, I.D. Soares,  L. Thorlacius and J. Zanelli.
We have benefited from partial financial support from CNPq, Faperj and Capes (Brazilian agencies).
\end{acknowledgements}


\end{document}